\documentclass[pre,
aps,superscriptaddress,floats,showpacs, 10pt]{revtex4-1}
\usepackage{graphicx}
\usepackage{dcolumn}
\usepackage{bm}
\usepackage{amssymb}
\usepackage{amsmath}
\usepackage{color}

\begin{document}

\title{On the breaking of a plasma wave in a thermal plasma:\\
II. Electromagnetic wave interaction with the breaking plasma wave}
\author{Sergei V. Bulanov}
\altaffiliation[Also at ]{A. M. Prokhorov Institute of General Physics of RAS, Moscow, Russia}
\affiliation{QuBS, Japan Atomic Energy Agency, 1-8-7 Umemidai, Kizugawa, Kyoto, 619-0215 Japan}
\author{Timur Zh. Esirkepov}
\affiliation{QuBS, Japan Atomic Energy Agency, 1-8-7 Umemidai, Kizugawa, Kyoto, 619-0215 Japan}
\author{Masaki Kando}
\affiliation{QuBS, Japan Atomic Energy Agency, 1-8-7 Umemidai, Kizugawa, Kyoto, 619-0215 Japan}
\author{James K. Koga}
\affiliation{QuBS, Japan Atomic Energy Agency, 1-8-7 Umemidai, Kizugawa, Kyoto, 619-0215 Japan}
\author{Alexander S. Pirozhkov}
\affiliation{QuBS, Japan Atomic Energy Agency, 1-8-7 Umemidai, Kizugawa, Kyoto, 619-0215 Japan}
\author{Tatsufumi Nakamura}
\affiliation{QuBS, Japan Atomic Energy Agency, 1-8-7 Umemidai, Kizugawa, Kyoto, 619-0215 Japan}
\author{Stepan S. Bulanov}
\altaffiliation[Also at ]{Institute of Theoretical and Experimental Physics, Moscow 117218, Russia}
\affiliation{University of California, Berkeley, CA 94720, USA}
\author{Carl B. Schroeder}
\affiliation{Lawrence Berkeley National Laboratory, Berkeley, California 94720, USA}
\author{Eric Esarey}
\affiliation{Lawrence Berkeley National Laboratory, Berkeley, California 94720, USA}
\author{Francesco Califano}
\affiliation{Physical Department, University of Pisa, Pisa 56127, Italy}
\author{Francesco Pegoraro}
\affiliation{Physical Department, University of Pisa, Pisa 56127, Italy}
\date{19/Apr/2012, 12:30, Japan time}

\begin{abstract}
The structure of the  density singularity formed in a  relativistically large amplitude plasma wave  close 
to the wavebreaking limit 
leads to a refraction coefficient which has a coordinate dependence with discontinuous derivatives. 
This results in a non-exponentially
small above-barrier reflection 
of an electromagnetic wave
interacting with the nonlinear plasma wave.
\end{abstract}
\maketitle

\section{Introduction}

In the first part of our paper \cite{Part-I}, extending an approach 
formulated in Ref. \cite{RCD} to the relativistic limit, we have studied systematically 
the structure of the  singularities formed in a  relativistically large amplitude plasma wave  close 
to the wavebreaking in a thermal plasma. We have shown that typically the electron density distribution
in the  breaking wave has a ``peakon" form with  a discontinuous coordinate dependence of its  first derivative,
similar to the profiles of nonlinear water waves \cite{Stokes, Whitham, peakon} 
{and  that in the above breaking limit the derivative becomes infinite}. 
This results in a finite reflectivity of an electromagnetic wave interacting with 
nonlinear plasma waves.
In particular, this is  an important property because  nonlinear Langmuir waves play a key role 
in the ``relativistic flying mirror" concept \cite{RFM, RFM1, RFM1a, RFM1b, RFM1c, RFM2}. 
In this concept, very
high density electron shells are formed in the nonlinear wake wave generated
by an ultrashort laser pulse propagating in an underdense plasma with   a 
speed close to  the  speed of light in vacuum. The shells act as  mirrors
flying with relativistic velocity. When they reflect a counterpropagating
electromagnetic pulse, the pulse   is compressed,  its frequency is upshifted  and its
intensity increased. It is the singularity in the electron density
distribution that  allows for a  high efficiency  in  the reflection of a portion of
the counterpropagating electromagnetic pulse. If the Langmuir wave is far
below the wave-breaking threshold, its reflectivity is exponentially small.
For  a nonlinear Langmuir wave the singularity formed in the electron density
breaks   the geometric optics approximation and leads to a reflection coefficient 
that is not exponentially small \cite{RFM, PAN}. 

In the present paper  we address the problem of the interaction of an electromagnetic wave with a 
nonlinear plasma wave which is of interest for the ``photon accelerator" concept \cite{WILKS} and for 
the ``relativistic flying mirror" paradigm \cite{RFM, RFM1, RFM1, RFM1a, RFM1b, RFM1c, RFM2}. 
We calculate the reflection coefficients of an
electromagnetic wave at the singularities of the electron
density in the most typical regimes of a strongly
nonlinear wave breaking in thermal plasmas.

\bigskip

\section{Electromagnetic wave reflection by the
electron density modulated in the breaking wave}

 As we have seen in the first part of our paper \cite{Part-I}, 
 in a strongly nonlinear wake wave the electron density is modulated 
and forms  thin shells (singularities or caustics in the  plasma flow) moving with 
 velocity $\beta_{\rm ph}$. 
In the Introduction, in a way of Refs. \cite{RFM, RFM1, RFM1a, RFM1b, RFM1c, RFM2},  we have discussed 
how a  counterpropagating electromagnetic wave
 can be partially reflected from these density shells which play the role of relativistic mirrors.
While in the case of a cold plasma the electron density at the singularity  tends to infinity 
(see Eq. (59) of Part I \cite{Part-I} and Refs. \cite{RFM1, PAN}),
in  a thermal plasma  the density is limited by  the expressions given by Eqs. 
(41) and (42) of Part I \cite{Part-I}. 
Although in this case the density profile is described 
by a  continuous  function of  the variable  $X$, its derivatives with respect to $X$ are discontinuous. 
This discontinuity results 
in the breaking of the  geometric optics approximation and leads to a 
reflectivity that is  not exponentially small.

In order to calculate the reflection coefficient, we consider the interaction of an
electromagnetic wave with the electron density shell formed  at the  breaking point  of a  Langmuir wave
in  a thermal plasma similarly to what has been 
 done in Refs. \cite{RFM, PAN}. The electromagnetic wave, described by the $z$ component 
 of the vector potential
$A_z(x,y,t)$, evolves according to the linearized wave equation
\begin{equation}
\partial_{tt} A_z-c^2 \left(\partial_{xx} A_z + \partial_{yy} A_z \right) +
\Omega_{pe}^2(x-v_{\rm ph}t)A_z=0,
\label{eq25-weq}
\end{equation}
where we have reverted to dimensional units and 
\begin{equation}
\Omega_{pe}^2(X)=\frac{4 \pi e^2}{m_e } 
\int_{-\infty} ^{+\infty} \frac{f_e(p) \mathrm{d}p}{\sqrt{1+(p/m_ec^2)^2}}.
\label{eq26a-Ompe}
\end{equation}
The last term in the l.h.s. of Eq. (\ref{eq25-weq}) is the $z-$component of the electric current density 
generated by the electromagnetic wave in a plasma with the electron distribution function $f_e(p)$. 
In the limit $\gamma_{\rm ph}\Delta p_0\ll 1$ for the electromagnetic wave frequency larger than
 the Langmuir frequency calculated for the maximal electron density, 
 $\omega \gg \omega_{\rm pe} (2 \gamma_{\rm ph}/\Delta p_0 )^{1/4}$, we can neglect the finite temperature effects 
on the electromagnetic wave dispersion, which have been analyzed in Ref. \cite{SILIN}, in the limit of 
homogeneous, stationary plasmas.
For the water-bag distribution function 
\begin{equation}
f_{e}(p,X)=n_{0} \theta(p-p_-(X))
\theta(p_+(X)-p)/\Delta p_0
\end{equation}
$\Omega_{pe}^2(X)$ takes the form
\begin{equation}
\Omega_{pe}^2(X)=\omega_{pe}^2 \frac{1}{ \Delta p_0} \ln \left(\frac{p_{+}(X)+
\sqrt{1+ p_{+}(X)^2}}{p_{-}(X)+\sqrt{1+ p_{-}(X)^2}} \right),
\label{eq26-Ompe}
\end{equation}
where $\omega_{pe}^2=4\pi n_0 e^2/m_e$, $p_{\pm}(X)$ and $\Delta p_0$ are 
now dimensionless (normalized on $m_e c$).

\begin{figure}[tbph]
\includegraphics[width=8cm,height=5cm]{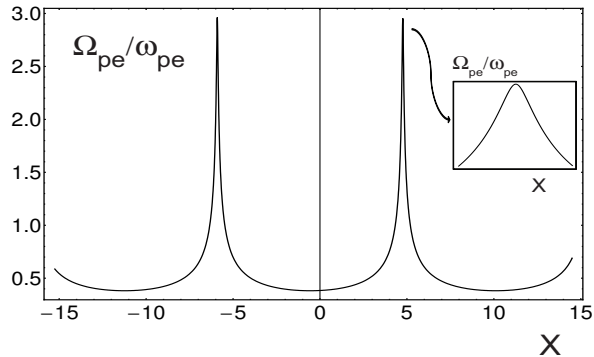}
\caption{Dependence of
the frequency ratio $\Omega_{pe}/\omega_{pe}$ on the coordinate $X$ for the parameters 
of a wakewave corresponding to $\Delta p_0=0.1$ and $E_{\rm max}=2.3$ at $X=15$ and 
for $\beta_{\rm ph}=0.992$. In the inset the ratio $\Omega_{pe} (X)/\omega_{pe}$ 
is shown in the vicinity of the maximum.}
\label{fig14}
\end{figure}

The wake wave modulates the electron density and temperature increasing them 
in the compression regions and decreasing 
them in the rarefaction regions. 
In Fig. \ref{fig14} we illustrate the dependence of $\Omega_{pe}(X)/\omega_{pe}$ 
on $X$
 for the parameters of a wakewave corresponding to $\Delta p_0=0.1$ and $E_{\rm max}=2.3$ at $X=15$ 
 and for $\beta_{\rm ph}=0.992$.

From Eqs. (35) and (59) of Part I \cite{Part-I} in the ultrarelativistic case, $\beta_{\rm ph}\approx 1$, 
using Eq. (\ref{eq26-Ompe}) we find for a relatively cold distribution such that
$p_{-}\ll 1$ that near the wavebreaking point
$\Omega_{pe}^2(X)$ is given by
\begin{equation}
\Omega_{pe}^2(X)\approx  \frac{\omega_{pe}^2}{\gamma_{\rm ph}} 
-\frac{\omega_{pe}^2 \sqrt{n_{\rm br} \gamma_{\rm ph}}}{\Delta p_0}|X|.
\label{eq25-weq}
\end{equation} 

The propagation of a sufficiently short electromagnetic wave packet 
 in the plasma with electron density 
modulated by the Langmuir wave can be described within the framework of the geometric optics approximation. 
The electromagnetic wave 
is represented as a particle (``photon") with coordinate  $x$ and momentum ${\bf k}$ (wave vector).
The interaction of  a ``photon" with a Langmuir wave that propagates with a relativistic phase velocity 
$v_{\rm ph}\approx c$ can be accompanied by a 
substantial frequency upshift called ``photon acceleration" \cite{WILKS, FA0, FA1, FA2}. 
Using the dispersion equation 
\begin{equation}
\omega(x,{\bf k};t)=\sqrt{k^2c^2+\Omega^2_{pe}(x-v_{\rm ph}t)},
\label{eq25-omxkt}
\end{equation}
where $k^2=k^2_{||}+k^2_{\perp}$ with $k_{||}$ and $k_{\perp}$  the wave vector components parallel 
and perpendicular to the propagation direction of the  Langmuir wave,
we obtain the "photon" Hamiltonian function which depends on the canonical variables $X=x-v_{\rm ph} t$ and $k_{||}$ 
(see Ref. \cite{TZE})
\begin{equation}
{\cal H}_{\rm photon}(X,k_{||})=\sqrt{k_{||}^2c^2+\Omega^2_{pe}(X)}-\beta_{\rm ph}k_{||}c.
\label{eq25-Ham-phtn}
\end{equation}
The transverse component of the wave vector is constant $k_{\perp}=k_{\perp,0}$ 
and  $k_{\perp,0}=0$ is assumed for the sake of simplicity.

\begin{figure}[tbph]
\includegraphics[width=9cm,height=6cm]{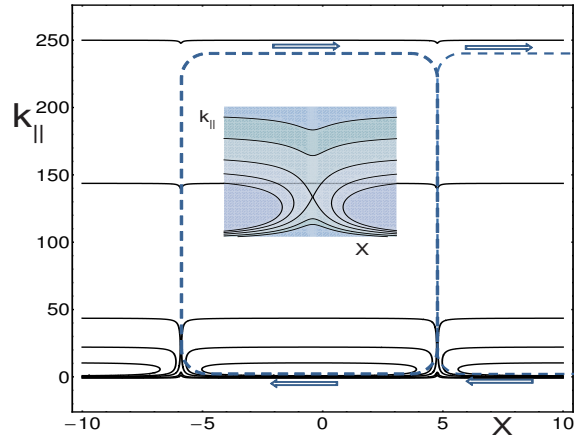}
\caption{Photon phase portrait  for the parameters of  a  nonlinear Langmuir wave corresponding to 
Fig. \ref{fig14}. The dashed line corresponds to the trajectory of photons that have appeared due to the over-barrier 
reflection at the crest of the breaking wave. 
In the inset the photon trajectories in the vicinity of the saddle point are shown.}
\label{fig15}
\end{figure}

The phase portrait of the photon for $\Omega_{pe}(X)$ given by Eq. (\ref{eq26-Ompe}) for the parameters corresponding to 
Fig. \ref{fig14} is shown in Fig. \ref{fig15}.

 Along  an orbit corresponding to  the value 
${\cal H}_{\rm photon}(X,k_{||})={\cal H}_{\rm photon}(X_0,k_{||,0})={\cal H}_{\rm photon,0}$ 
of  the Hamiltonian (\ref{eq25-Ham-phtn}) 
 the photon frequency is given by 
\begin{equation}
\omega=
\gamma_{\rm ph}^2 {\cal H}_{\rm photon,0}
\left[1 \pm \beta_{\rm ph}
\sqrt{1-\frac{\Omega_{pe}^2(X)}{{\cal H}_{\rm photon,0} \gamma_{\rm ph}^2}}\right].
\label{eq25-omega-phtn}
\end{equation}

Photons, for which ${\cal H}_{\rm photon,0}<{\rm max}\{\Omega_{pe}/\gamma_{\rm ph}\}$ 
are trapped inside the region 
encircled by the separatrix.
Along the orbit their frequency changes from $\omega_{{\rm max}}$ and $\omega_{{\rm min}}$ corresponding 
to the plus and minus 
signs in the r.h.s. of Eq. (\ref{eq25-omega-phtn}) at the minimum of $\Omega_{pe}(X)$.

Photons with ${\cal H}_{\rm photon,0}>{\rm max}\{\Omega_{pe}/\gamma_{\rm ph}\}$ 
are not trapped and for them 
the sign in the r.h.s. of Eq. (\ref{eq25-omega-phtn}) does not change. 
For trajectories far above the separatrix 
the photon frequency variations are relatively weak. 
However a sufficiently
strong wakefield can reflect a counterpropagating
photon, $k_{||,0} < 0$ due to above-barrier reflection 
(this trajectory is shown in Fig. \ref{fig15} by  a dashed line). 
Such a photon acquires  a frequency 
\begin{equation}
\omega = \omega_0 \frac{1+\beta_{\rm ph}}{1-\beta_{\rm ph}}
\end{equation}
according to the Einstein formula for the frequency 
of the electromagnetic wave reflected by  a relativistic mirror \cite{AEin}.
The geometric optics approximation
fails when the wakefield is close to wave breaking and this provides the appropriate 
conditions for  a not exponentially weak wave scattering.

\section{Above-barrier scattering of an electromagnetic wave at the crest of the breaking wake wave}

In order to find the reflectivity of the nonlinear wake wave we make a Lorentz transformation 
to the frame of reference
 moving with the phase velocity of the Langmuir wave. 
 In the  boosted frame, Eq.   (\ref{eq25-weq}) 
for the electromagnetic wave interacting with the nonlinear Langmuir wave 
can be written as 
\begin{equation}
\frac{d^2 a(\zeta)}{d \zeta^2}+q^2(\zeta) a(\zeta)=0
\label{eq33-aX}
\end{equation}
with
\begin{equation}
 a(\zeta)= \frac{e A_z}{m_e c^2} \exp{\left[-i(\omega' t'-k_y y)\right]}
\label{eq34-az}
\end{equation}
and in the neighbourhood of the breaking point $q^2(\zeta)$ can be written as
\begin{equation}
 q^2(\zeta)= s^2 + g_{-1}|\zeta|.
\label{eq35-q2}
\end{equation}
Here
\begin{equation}
s^2=\frac{\omega'^2}{c^2}-k_y^2-\frac{\omega_{pe}^2}{c^2\gamma_{\rm ph}},
\end{equation}
 and $\zeta=X\gamma_{\rm ph}$, $t'$, $k'$, $\omega'$ are the coordinate 
and time and the wave number and frequency in the boosted frame of reference.
The coefficient $g_{-1}$ is equal to
\begin{equation}
g_{-1}=\frac{\omega_{pe}^2 \sqrt{n_{\rm br}}}{ \Delta p_0 c^2 \sqrt{\gamma_{\rm ph}}}.
\end{equation}
\begin{figure}[tbph]
\includegraphics[width=9cm,height=6cm]{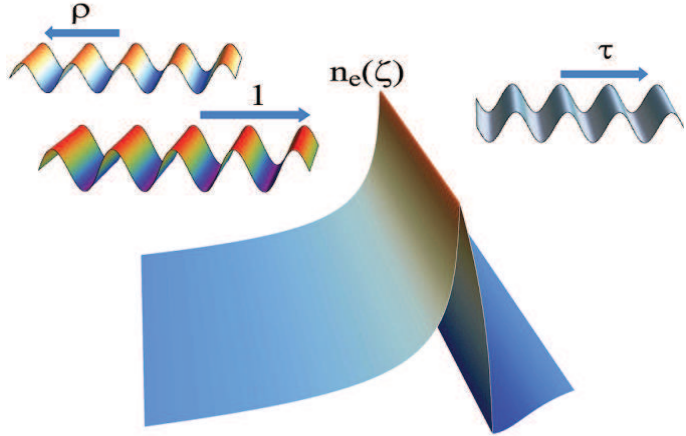}
\caption{Scattering geometry.}
\label{fig16}
\end{figure}

We seek for the solution to the above-barrier scattering problem for Eq. (\ref{eq33-aX}) writing 
its  solution 
 in the form (see Refs. \cite{PAN, POKR, M-H})
\begin{equation}
a(\zeta)=\frac{1}{\sqrt{q(\zeta)}} \left[b_+ \exp{\left(i W(\zeta)\right)} 
+b_- \exp{\left(-i W(\zeta)\right)}\right],
\label{eq40-sol}
\end{equation}
where the phase integral is defined as
\begin{equation}
W(\zeta)=\int_0^{\zeta} q(\zeta') d \zeta'.
\label{eq40-W}
\end{equation}

The above-barrier scattering geometry is illustrated in Fig. \ref{fig16}.
For constant $b_+$ and $b_-$  Eq. (\ref{eq40-sol}) corresponds to  the ``WKB" solution \cite{FROFRO}.
 In the following the  coefficients $b_+$ and $b_-$ are considered as functions of $W$ instead of $\zeta$,
because, as explained in Ref. \cite{BERRY}, the mapping between $W$ and $\zeta$ 
given by Eq. (\ref{eq40-W}) 
is one-to-one on the real axis.
Far from the breaking point, i.e. formally for $\zeta \to \pm \infty$
the function $q^2(\zeta) \to s^2 $ reduces to a constant and the solutions (\ref{eq40-sol}) 
are exact, so that $\ b_{\pm}\to  const$ as $W \to \pm \infty$.

In other words, the boundary conditions at $\zeta \to \pm \infty$ are
\begin{equation}
\begin{array}{c}
b_+(+\infty)=1, \qquad b_-(+\infty)=\rho,\\
b_+(-\infty)=0, \qquad b_-(-\infty)=\tau.
\end{array}
\label{eq44-bcond}
\end{equation}

Since in the representation (\ref{eq40-sol}), 
the single unknown function $a(\zeta)$ has been replaced by the 
two unknown functions $b_{\pm}(\zeta)$, a subsidiary condition is necessary. 
 We shall impose the condition
\begin{equation}
\frac{d a}{d \zeta}=i\sqrt{q(\zeta)} \left(b_+ e^{i W(\zeta)} 
-b_- e^{-i W(\zeta)}\right).
\label{eq43-subs}
\end{equation}
Differentiating Eq.  (\ref{eq40-sol})  with respect to $\zeta$ and taking into account 
the constraint (\ref{eq43-subs}), we find
\begin{equation}
\frac{d b_{+}}{d \zeta}e^{i W(\zeta)}+\frac{d b_{-}}{d \zeta}e^{-i W(\zeta)}=\frac{d \ln{\sqrt{q}}}{d \zeta}
\left(b_{+} e^{i W(\zeta)}+b_{+} e^{-i W(\zeta)}\right), 
\label{eq45-bpm1}
\end{equation}
while differentiating Eq. (\ref{eq43-subs}) with respect to $\zeta$ and substituting  $d^2 a/d \zeta^2$ 
into Eq.  (\ref{eq33-aX}) yields
\begin{equation}
\frac{d b_{+}}{d \zeta}e^{i W(\zeta)}-\frac{d b_{-}}{d \zeta}e^{-i W(\zeta)}=\frac{d \ln{\sqrt{q}}}{d \zeta}
\left(b_{-} e^{-i W(\zeta)}-b_{+} e^{i W(\zeta)}\right).
\label{eq45-bpm2}
\end{equation}
The system of Eqs. (\ref{eq45-bpm1}) and (\ref{eq45-bpm2}) is equivalent to Eq. (\ref{eq33-aX}). 
It can be rewritten  in the form
\begin{equation}
\frac{d }{d W}
\left(
\begin{array}{c}
b_+\\
b_-
\end{array}
\right)
=
\left(
\begin{array}{cc}
0&S(W) e^{2iW}\\
S(W) e^{-2iW}&0
\end{array}
\right)
\left(\begin{array}{c}
b_+\\
b_-
\end{array}
\right),
\label{eq44-b}
\end{equation}
with
\begin{equation}
S(W)=\frac{1}{2} \frac{d}{dW} \ln{q(\zeta(W))}. 
\label{eq45-SW}
\end{equation}
For $q(\zeta)$ given by Eq. (\ref{eq35-q2}) we have
\begin{equation}
W(\zeta)=\frac{2}{3g_{-1}}\left[\left(s^2+g_{-1}|\zeta|\right)^{3/2}-s^3\right] \, {\rm sign}(\zeta), 
\label{eq45-SW11}
\end{equation}
where ${\rm sign}(\zeta)=-1$ if $\zeta<0$ and ${\rm sign}(\zeta)=1$ for $\zeta>0$,
and 
\begin{equation}
q(W)=\left(\frac{3g_{-1}}{2} W {\rm sign}(\zeta) +s^3\right)^{1/3}
\label{eq45-SW12}
\end{equation}
so that
\begin{equation}
S(W)=\frac{g_{-1}}{4 q^3(\zeta(W))}{\rm sign}(\zeta). 
\label{eq45-SW1}
\end{equation}
It follows that $S(W)$ is discontinuous at $\zeta\to 0$ ($W \to 0$)
\begin{equation}
S(W=0)=\frac{g_{-1}}{4 s^3}{\rm sign}(\zeta). 
\label{eq45-SD}
\end{equation}

Integrating both sides of Eq. (\ref{eq44-b}) and using the above formulated boundary conditions 
for $b_{\pm}(\pm \infty)$,
we can obtain the reflection coefficient $\rho$ in the form of the infinite series \cite{BERMOUN}
\begin{equation}
\rho=-\displaystyle{\sum^{\infty}_{m=0}(-1)^m\int^{+\infty}_{-\infty} dW_0 S(W_0) e^{2iW_0}}
\displaystyle{
\prod_{n=1}^{m}\int_{-\infty}^{W_n-1} dV_n S(V_n)e^{-2iV_n}}
\displaystyle{
\int^{+\infty}_{V_n}dW_{n}S(W_n)e^{2iW_n}},
\label{eq46-rho}
\end{equation}
with the product equal to unity for $m=0$.

The function $ q(\zeta)$ defined by Eq. (\ref{eq35-q2}) 
has a  discountinous  first derivative at $\zeta=0$.
In the vicinity of the singularity point 
it can be represented in the form $q(\zeta)\approx q_0+q_1 |\zeta|$ with $q_0=s$ and
$q_1=g_{-1}/2s$.
Expanding $W(\zeta)$ and $S(W)$ in powers of $\zeta$ 
and substituting them into Eq. (\ref{eq46-rho}) we can find (see Eq. (27) of Ref. \cite{BERRY})
that the first term yields the dominant contribution to the reflection coefficient, with the result
\begin{equation}
\rho_{-1}\approx \frac{-i q_1}{s q_0}=\frac{-i g_{-1}}{4 s^3}. 
\label{eq47-rho}
\end{equation}
and
\begin{equation}
R_{-1}=|\rho_{-1}|^2=  g_{-1}^2\, \frac{1}{s^6}.
\label{eq-R-1}
\end{equation} 
Applicability of the WKB theory implies that $g_{-1}\ll s^3$.

Similarly (see also Ref. \cite{PAN}) we can find the reflection coefficient at the electron density singularity
formed in the above breaking regime discussed in Part I \cite{Part-I}. In this case the electron density 
distribution is given by Eq. (106) of Part I. Using this relationship we obtain 
\begin{equation}
\rho_{\left(\frac{1}{2},\frac{1}{2}\right)}= \frac{-4 i g_{\left(\frac{1}{2},\frac{1}{2}\right)}}{s }
\int_{-\infty}^{+\infty}{\exp{[2 i s \zeta]}\left(\theta (\zeta) \sqrt{\zeta}
-\theta (\zeta-\Delta \zeta) \sqrt{\zeta-\Delta \zeta}\right) d\zeta}, 
\label{eq-rho-half-half}
\end{equation}
where 
\begin{equation}
g_{\left(\frac{1}{2},\frac{1}{2}\right)}=k_p^{3/2}\gamma_{\rm ph}^{3/2}\frac{\sqrt{2 e E_{\max} m_e c}}
{\Delta p_0} ,
\end{equation}
$k_p = c/  \omega_{pe}$
 and 
$\Delta \zeta = \Delta p_0/e E_{\max}$. Calculating the integral (\ref{eq-rho-half-half}) we find
\begin{equation}
\rho_{\left(\frac{1}{2},\frac{1}{2}\right)}=g_{\left(\frac{1}{2},\frac{1}{2}\right)} 
(1+i)\sqrt{2\pi}\,\frac{\exp{\left( is \Delta \zeta\right)}\sin{(s \Delta \zeta)}}{s^{5/2}}.
\label{eq-rho-half-half-2}
\end{equation}
Consequently, we write
\begin{equation}
R_{\left(\frac{1}{2},\frac{1}{2}\right)}=|\rho_{\left(\frac{1}{2},\frac{1}{2}\right)}|^2
=  g_{\left(\frac{1}{2},\frac{1}{2}\right)}^2\,4\pi \, \frac{{\rm sin}^2{(s \Delta \zeta)}}{s^5}.
\label{eq-R-half-half}
\end{equation}
From Eqs. (\ref{eq-R-1}) and (\ref{eq-R-half-half}) we can see that in thermal plasmas 
the reflection coefficient is $s\gg1$ times larger in the above breaking regime than for a wake wave 
approaching the wavebreaking threshold.

Generalizing Eqs. (106) and (108) of Part I \cite{Part-I}, we can write the electron 
density dependence on the coordinate $\zeta$ in the form 
\begin{equation} 
\label{eq:w-b-n_e-even}
n_e(\zeta) \sim 
\frac{2 n_0}{\Delta \zeta}
\left[ 
\theta \left(\zeta_{+}\right) \left(\zeta_{+}\right)^{1/m}
-
\theta \left(\zeta_{-}\right) \left(\zeta_{-}\right)^{1/m}
\right]
\end{equation}
with $\zeta_{\pm}=\zeta\mp \Delta \zeta/2$ and  $m$  an even number, and 
\begin{equation} 
\label{eq:w-b-n_e-odd}
n_e(\zeta) \sim 
\frac{n_0}{\Delta \zeta}
\left[ 
\theta \left(\zeta_{+}\right) \left(\zeta_{+}\right)^{1/m}
+\theta \left(-\zeta_{-}\right) \left(-\zeta_{-}\right)^{1/m}
-
\theta \left(\zeta_{-}\right) \left(\zeta_{-}\right)^{1/m}
-\theta \left(-\zeta_{+}\right) \left(-\zeta_{+}\right)^{1/m}
\right]
\end{equation} 
for $m$  an odd number.

It is easy to show that for the reflection coefficient, 
$R_{\left(\frac{1}{m},\frac{1}{m}\right)}=|\rho_{\left(\frac{1}{m},\frac{1}{m}\right)}|^2$, we have 
\begin{equation}
R_{\left(\frac{1}{m},\frac{1}{m}\right)}=g_{\left(\frac{1}{m},\frac{1}{m}\right)}^2\,
\left|{4 (-i s)^{-1/m} \Gamma \left(1 + \frac{1}{m}\right) \frac{\sin (s \Delta \zeta) }{s \Delta \zeta}}\right|^2,
\label{eq-R-odd}
\end{equation}
if $m$ is  even, and 
\begin{equation}
R_{\left(\frac{1}{m},\frac{1}{m}\right)}=g_{\left(\frac{1}{m},\frac{1}{m}\right)}^2\,
\left|{2 (1 + (-1)^{1/m}) (-is)^{-1/m}
  \Gamma\left(1 + \frac{1}{m}\right) \frac{\sin (s \Delta \zeta )}{s \Delta \zeta}}\right|^2,
\label{eq-R-even}
\end{equation}
if  $m$ is odd.

Comparing Eqs (\ref{eq-R-half-half}, \ref{eq-R-odd}, \ref{eq-R-even}) for the reflection coefficient with the 
corresponding coefficients obtained in Ref. \cite{PAN}, we find that the effects of  a finate temperature 
enter Eqs. (\ref{eq-R-half-half}, \ref{eq-R-odd}, \ref{eq-R-even})
as a form-factor $|{\sin (s \Delta \zeta )}/{s \Delta \zeta}|^2$. In the limit $\Delta p_0 \to 0$  
this form factor 
tends to unity  while for $s \Delta \zeta \gg 1$ decreases as $\approx 1/(s \Delta \zeta)^2$.

Since the frequency, $\omega_r$, and the number of reflected photons, $N_r$, are related to that 
 incident on the relativistic mirror $\omega_0$ and $N_S$ as 
$\omega_r=\omega_0(1+\beta_{\rm ph})/(1-\beta_{\rm ph})\approx \omega_0 4 \gamma_{\rm ph}^2$ and $N_r=R N_S$,
the energy of the relected photon beam is given by ${\cal E}_r \approx {\cal E}_S 4 \gamma_{\rm ph}^2 R$,
where ${\cal E}_S$ is the energy of the laser pulse incident on the mirror. Comparing ${\cal E}_r$ with 
the energy of the electrons in the first period of the wake wave (e.g. see \cite{ESL1}), ${\cal E}_e \approx
{\cal E}_{\rm las,d} (\omega_{\rm pe}/\omega_0)^2$, where ${\cal E}_{\rm las,d}$ is the laser driver energy, we
find that the photon back reaction (the ponderomotive pressure) on the wake wave can be neglected provided
 ${\cal E}_S \ll {\cal E}_{\rm las,d}/4 \gamma_{\rm ph}^4 R$. As a typical reflection coefficient value we
 can take $R\approx 1/ \gamma_{\rm ph}^4$ (see Refs.\cite{RFM1b, PAN}) and obtain 
 the condition of relative weakness of the incident laser pulse ${\cal E}_S \leq {\cal E}_{\rm las,d}$.
 As we see owing to the weakness of the photon-wakewave interaction the incident pulse energy 
 can be of the order of that in the driver laser pulse.

\section{Discussions and Conclusions}

In the first Part of our paper \cite{Part-I} we  found the structure
of  the typical singularities that  appear  in the electron density
during the wave breaking in a thermal plasma. 
 The singularity in the electron
density, moving along with the wake wave excited by a high intensity ultra-short pulse laser, 
can act as a flying
relativistic mirror for counterpropagating electromagnetic
radiation, leading to  coherent reflection accompanied by the
 upshift of the  radiation frequency. This process implies finite (not exponentially small) 
reflectivity at the electron density singularities. This is 
 provided by the structure of the  singularity formed in a  
relativistically large amplitude plasma wave  close 
to the wavebreaking limit  that leads to  a  refraction coefficient  
with discontinuous coordinate  derivatives. 
We found the  reflection coefficients of an
electromagnetic wave at the singularities of the electron
density in the most typical regimes of strongly
nonlinear wave breaking in thermal plasmas. The efficiency of the photon reflection
can be substantially increased by using the above breaking limit regimes  which lead to
the formation of   high-order singularities.

\medskip
\begin{acknowledgments}
We acknowledge support of this work from We acknowledge the support from the MEXT of Japan,
Grant-in-Aid for Scientific Research, 23740413, and Grant-in-Aid 
for Young Scientists 21740302 from MEXT. We appreciate support from the NSF under Grant No. PHY-0935197 and 
the Office of Science of the US DOE under Contract No. DE-AC02-05CH11231.

\end{acknowledgments}

\end{document}